\documentclass[%
 reprint,
 amsmath,amssymb,
 aps,
]{revtex4-2}

\usepackage{graphicx}
\usepackage{dcolumn}
\usepackage{bm}

\newcommand{\del}{\partial}
\begin{document}

\preprint{APS/123-QED}

\title{Quantum Back Flow Across a  Black Hole Horizon in a Toy Model Approach}

\author{Dripto Biswas}\thanks{dripto.biswas@niser.ac.in}
\affiliation{
School of Physical Sciences\\
 National Institute of Science Education and Research \\
 Bhuabneswar, India
}%

\author{Subir Ghosh}
\affiliation{%
 Physics and Applied Mathematics Unit\\
Indian Statistical Institute\\
Kolkata, India
}%

\date{\today}

\begin{abstract}
Quantum Back Flow (QBF), discovered quite a few years back, is a generic purely quantum phenomenon, in which the probability of finding  a particle in a direction is non-zero (and increasing for a certain period of time) even when the particle has with certainty a velocity in the {\it opposite}   direction.  In this paper, we study  QBF of a quantum particle across the event horizon of a  Schwarzschild Black Hole. In a toy model approach, we consider a  superposition of two ingoing solutions  and observe the probability density and  probability current. We explicitly demonstrate a non-vanishing quantum back flow in a small region around the event horizon. This is in contrast to the classical black hole picture, that once an excitation crosses the horizon, it is lost forever from the outside world. Deeper implications of this phenomenon are speculated. We also study quantum backflow for another spacetime with horizon, the Rindler spacetime, where the phenomenon can be studied only within the Rindler wedge.
\end{abstract}

\maketitle

\section {Introduction} Consider a particle moving along, say, the $x$-axis having with certainty a positive velocity, along right hand direction with respect to some arbitrary origin $x=0$. For a classical particle obviously the chance that it will be detected at $x>0$ will increase with time. However, for a quantum particle, on the contrary, for certain wave functions, the probability of the particle's position at $x<0$ actually increases with time, for a finite period of time. This  generic  counter-intuitive phenomenon is the well known Quantum Back Flow (QBF). It can occur for   a free particle  or one in presence of a potential. Even if the particle's  wave function is centered around $x=0$ and is constructed out of only positive momenta, it can still possess non-vanishing probability of being observed at  $x < 0$ and the probability can even increase for finite periods of time. Thus, the quantum-mechanical current at the origin can be negative with the probability flowing {\it in opposite direction to the momenta}. In this paper, for the first time, we study QBF of ingoing modes crossing the Black Hole (BH) event horizon. Obviously this is contrary to conventional BH physics wisdom that nothing can come out of the BH horizon. It can have    deep impact in fundamental BH physics. It needs to be mentioned that QBF is a generic quantum phenomenon and in the QBF effects studied so far (please see the references in our paper [1-21]) the presence of a spacetime horizon (as in BH) does not arise.

 QBF was  revealed by Allcock's work \cite{alc} on the arrival time problem in quantum theory (see also \cite{mugg,muga}). It also appeared in the time operator construction in the Heisenberg time-energy uncertainty relation \cite{kij} and decay of quasistable quantum systems where the small inward current  was interpreted as QBF \cite{decay}. Bracken and Melloy \cite{bm}  showed that, (i) the total amount of QBF is bounded (by a dimensionless number computed
numerically $\approx  0.04$) and (ii) the bound has a universal character, independent of the time duration, particle mass and $\hbar $. A subtle   limiting procedure proves  vanishing of the effect in classical framework \cite{years}. A numerical study \cite{penz} revealed a structure of the wave function and the corresponding approximate analytic form of the wave function appeared in \cite{intro,years,sol} which yields a  modest backflow. Its relativistic extension was studied in  \cite{rel,ash,qpot}.  In our case, the BH horizon generates an effective potential and the impact of a linear potential was investigated in \cite{qpot}. Connections between QBF and superoscillations was noted in \cite{ber}. A non-zero probability flux in classically forbidden regions, in the expansion of a wave packet in free space was discussed in \cite{gus1}. (For a recent brief review see \cite{intro}.) In a  recent paper \cite{expt1} direct evidence of QBF in analogue optical system has been reported. Other  experimental  proposals have been  utilising Bose-Einstein condensates \cite{29,30}, using solutions  with both positive and negative momenta  \cite{mil,expt} and a recent suggestion of moving  the particle  in a ring \cite{gus2}. We now come to our main concern - possible relevance of QBF across a Black Hole (BH) horizon.

In classical General Relativity, nothing can come out of the region behind the event horizon of a BH (that surrounds the spacetime singularity) and reach an observer at a large distance. This idea was overthrown in a quantum framework whereby BHs can emit Hawking radiation which carries information about the mass, charge and angular momentum of the BH. So far this is the only possible mechanism  of information (in the form of only mass, angular momentum, and charge of the BH) exchange between the BH and outside world. In this perspective QBF might have significance in providing another channel of information exchange between BH and outside observer. However, further discussion on  this is outside  the scope of the present work where we have established only the presence of QBF across BH horizon.

Hawking's seminal work \cite{haw,haw2} showed that a BH can radiate  with a characteristic (Hawking) temperature  and ushered in BH thermodynamics, thus formalising the BH area-entropy connection  \cite{bek}. However, this originated  the so called information paradox -  BH evaporation leaves only thermal radiation with   a temperature that depends just on a few macroscopic BH parameters and no information about the BH constituents, transforming via   non-unitary evolution,  from a pure state to a mixed state.  To keep unitarity intact,  BH entropy has to follow the Page curve \cite{page} and only recently it is becoming clear \cite{info,info2,info3} how quantum entanglement can remove the  information paradox. An alternative approach, for modelling a quantum BH,  advocated by t'Hooft \cite{th1,th2}, hinges on applying unitary  quantum mechanics at the BH horizon with the BH satisfying the Schrodinger equation.  In this framework,  scattering on the black hole horizon in a partial wave basis is  studied in \cite{th3,th4}. We, on the other hand, apply the Schrodinger equation to the particles on the BH horizon. In \cite{m12,m13},   physical boundary conditions for the quantum particle wave equation at BH horizon reveal  exponentially damped or enhanced solutions suggesting that particles instead of  crossing the BH horizon,  are absorbed or reflected by it.

In this perspective, QBF across BH horizon assumes significance, leading to open questions: can the back flow be directly interpreted as particles or at least can it transfer information across the horizon from the inside in  some form of  correlation between the ingoing wave and its QBF component?  Although we will revisit this issue at the concluding section, we address a point that is bound to come up, {\it i.e.} is there any connection between QBF and the celebrated Hawking effect, leading to Hawking radiation. Let us just emphasize that these two are entirely different phenomena: the fundamental differences being that Hawking radiation is a quantum field theoretic phenomenon derived in a  semi-classical framework where presence of the event horizon is crucial, whereas QBF is a quantum mechanical process and in the examples studied so far, the presence of an event horizon did not arise. The outcome of Hawking effect is that the radiation arises at the expense of the black hole mass and in the process the black hole might disappear completely. On the other hand the QBF we are looking at, external particles crossing in to the black hole are considered where part of this incoming flux can remain outside the horizon and the question of black hole evaporation as a result of QBF does not arise. An underlying motivation of our work is the possibility that QBF might have some correlation with the major portion of ingoing flux.

We consider a simplified scenario,  following earlier works \cite{bm}, and study QBF pertaining to a  superposition of two solutions of the  Schrodinger equation, near the BH horizon. (We leave the more realistic   problem of using wave packets for a future  publication.) Since our model is time dependent but {\it stationary}, with no temporal fall-off, we cannot use the conventional quantitative measures for QBF and can only establish  its presence  conclusively. We believe that these technical problems (such as wave packet construction), can be addressed straightforwardly in a more detailed analysis.

We have also briefly studied QBF for Rindler spacetimes, for which the metric allows a horizon. However the situation is qualitatively    different in this case since we work in Rindler coordinates where the constant acceleration in flat Minkowski spacetime induces an effective curvature (see for example \cite{rind}). We use the same procedure as in the Schwarzschild case. However, the question of QBF \textit{across} the horizon is not relevant here since Rindler coordinates are not defined beyond the Rindler wedge. We have computed the non-vanishing QBF within the Rindler wedge in Sec. \ref{sec:rind}.

The paper is organized as follows: in Section II, the basic formalism of constructing the Schrodinger equation for the curved spacetime Hamiltonian is elaborated. Further it is applied in the case of the Schwarzschild metric and the formal, analytic structure of the wave function is derived. In Section III, the observables relevant to QBF are introduced and the numerical analysis along with graphical representation of the main results are provided. This section constitutes our principal findings. In Section IV, we present the QBF results for Rindler spacetime along with relevant plots. Section V consists of a summary of our findings along with open future  challenges. An Appendix is provided at the end showing some intermediate computational steps.

 \section {Setting up the Schrodinger equation:}   From \cite{hertz} we write down the Hamiltonian in a curved background, where the BH metric in Cartesian coordinates reads,
\begin{equation}
\begin{split}
g^{00}=\frac{1}{U}&;~~g^{ij}=-\left[\eta^{ij}+(U-1)\frac{x^ix^j}{r^2}\right];\\
&r^2=(x_1)^2+(x_2)^2+(x_3)^2
\end{split}
\label{c1}
\end{equation}
\begin{equation}
g_{00}=U;~~g_{ij}=-\left[\eta_{ij}+(\frac{1}{U}-1)\frac{x_ix_j}{r^2}\right]
\label{c2}
\end{equation}
with ${\sqrt{-g}}=1$ and   $U=(1-\lambda/r)$ for the Schwarzschild metric. Following \cite{hertz}, $H$ is given by,
\begin{align}
\begin{split}
&H=\left[\frac{1}{\sqrt{-g}g^{00}}{\cal A}+\sqrt{-g}g^{ij}\partial_i\partial_j\frac{1}{{\cal A}}+m^2\sqrt{-g}\frac{1}{{\cal A}}\right] \\
&=U{\cal A}-\left[\eta^{ij}+(U-1)\frac{x^ix^j}{r^2}\right]\partial_i\partial_j\frac{1}{\cal A}+ m^2\frac{1}{\cal A},
\end{split}
\label{c3}
\end{align}
where, ${\cal A} =\sqrt{-\nabla^2+m^2}=\sqrt{-\eta_{ij}\partial^i\partial^j+m^2} $.  We reduce $H$ to the form,
\begin{equation}
\begin{split}
H=H_0&+V;~H_0={\cal A},\\
V=-\frac{\lambda}{r}{\cal A}-\nabla^2 \frac{1}{\cal A}&+\frac{\lambda}{r^3}x^ix^j\partial_i\partial_j \frac{1}{\cal A} +m^2\frac{1}{\cal A} 
\end{split}
\label{c5}
\end{equation}
leading to the Schrodinger equation, $i\partial_t \psi -H_0\psi = -V\psi$,
\begin{equation}
\begin{split}
 i\partial_t \psi -{\cal A}\psi &=-\left[-\frac{\lambda}{r}{\cal A}-\nabla^2 \frac{1}{\cal A}+\right.\\
 &\left.\frac{\lambda}{r^3}x^ix^j\partial_i\partial_j \frac{1}{\cal A} +m^2\frac{1}{\cal A}\right]\psi .
\end{split}
\label{c6}
\end{equation}
Recall that QBF in an external potential was studied in \cite{qpot}.
Putting back the fundamental constants, the above equation takes the form:
\vspace{0.85cm}
\begin{widetext}
\begin{align}
\begin{split}
i\hbar\partial_t \psi -\sqrt{-c^2\hbar^2\nabla^2 +c^4m^2}~&\psi 
=-\left[\frac{2Gm}{c^2r}\sqrt{-c^2\hbar^2\nabla^2+c^4m^2}
-\hbar^2\nabla^2\frac{1}{\sqrt{-c^2\hbar^2\nabla^2 +c^4m^2}} \right.\\ +
&\left.\frac{2\hbar^2GM}{r^3}x^ix^j\partial_i\partial_j\frac{1}{\sqrt{-c^2\hbar^2\nabla^2 +c^4m^2}}+m^2c^4\frac{1}{\sqrt{-c^2\hbar^2\nabla^2 +c^4m^2}}\right]\psi .
\end{split}
\label{c61}
\end{align}
\end{widetext}
\vspace{-0.6cm}
We consider ingoing solutions along the $X$-axis as shown in Fig. \ref{fig:diagram}, with a static observer outside the horizon ($r=\lambda $) on the $X$-axis.  Let us isolate the free plane wave part $\psi_u $ from $\psi$,  $\psi =\psi_u +\psi_s$ with  $\psi_u=Ae^{i(-Et + \vec k.\vec x)}$ and subsequently
\begin{equation}
[i\partial_t \psi_u -H_0\psi_u +V\psi_u] +  [i\partial_t \psi_s -H_0\psi_s +V\psi_s]
=0 .
\label{c7}
\end{equation}
In a perturbative framework we consider exact solutions of the free equation $i\partial_t \psi_u -H_0\psi_u =0$ and subsequently calculate the first order (in $V$) correction $\psi_s$ from the equation $[i\partial_t \psi_s -H_0\psi_s +V\psi_s] +V\psi_u =0$. Now $V\psi_u$ acts as a source in a Greens function scheme and we drop $V\psi_s$ since it will provide a correction effectively of $O(V^2)$ whereas  we are interested in  $O(V)$ corrections only. (We refer to the formalism used by  Allali and Hertzberg \cite{hertz}. 

Using the free particle dispersion relation
\begin{equation}
E_k-\sqrt{\hbar^2 c^2 k^2+m^2 c^4} =0
\label{c8}
\end{equation}
the equation for $\psi_s $ is given by
\begin{equation}
 [i\partial_t \psi_s -H_0\psi_s +V\psi_s] + V\psi_u =0
\label{c8}
\end{equation}
Considering a  first order potential correction for $\psi_s$, we drop the $V\psi_s$ term to obtain,
\begin{equation}
 i\partial_t \psi_s -H_0\psi_s  + V\psi_u =0 .
\label{c9}
\end{equation}
The computational details are provided in the supplemental material. We exploit Green's function techniques to first order, assuming the BH field to be weak near the horizon ($\sim$ a supermassive black hole whose Schwarzschild radius exceeds  its physical radius). The scattering part of $\psi$ is, therefore,
\begin{equation}
\begin{split}
\psi^{(k)}_s(x,t)=&\frac{-AE_k}{2\pi}e^{-iE_kt}\left[E_k\frac{e^{ik|x|}}{|x|}\int d^3x' \right.\\
\left(1-\frac{\lambda}{|x'|}\right) e^{i\vec q.\vec x'} &- \frac{\lambda \hbar^2 c^2}{E_k}k_i k_j \frac{e^{ik|x|}}{|x|}\int d^3x' \left. \frac{x^{'i}x^{'j}}{|x'|^3} e^{i\vec q.\vec x'}\right],
\end{split}
\label{c15}
\end{equation}
where, $A$ is an arbitrary constant and $\lambda = 2 GM/c^2$. Here, $\vec q = \vec k - k \hat{x}$. One can interpret $\vec q$ as a measure of the `scattering' as observed along $\hat{x}$. For e.g. if $\vec k = k \hat{x}$ (the wave is directed \textit{away} from the BH horizon), we expect no `scattering' to be observed. Indeed in this case, $\vec q = 0$.

We will work with the form of  $\psi^{(\vec k)} =\psi^{(\vec k)}_u +\psi^{(\vec k)}_s$, where, $\psi^{(\vec k)}_u=Ae^{i(-Et + \vec k.\vec x)}$ and $\psi^{(\vec k)}_s $ is given by (\ref{c15}).

{\bf Working form of the wave function $\psi^{(\vec k)}(\vec x,t)$:}
Expressing (\ref{c15}) in the following form
\begin{equation}\label{psis}
\begin{split}
    \psi_s^{(\vec k)}(\vec x,t) =-\frac{A E_k}{2\pi}&e^{-iE_k t} \frac{e^{i k |\vec x|}}{|\vec x|}[E_k \left(F_1(\vec q) - \lambda F_2(\vec q)\right) \\
     + &\frac{\lambda \hbar^2 c^2}{E_k}k_i k_j \del_{q_i} \del_{q_j} F_3(\vec q)],
\end{split}
\end{equation}
 reveals  three Fourier transforms,
\begin{equation}
\begin{split}
F_1(q)&=\int d^3x'~1~ e^{i\vec q.\vec x'}=(2\pi)^3\delta (\vec q), \\
F_2(q)=\int d^3x'&~\frac{1}{|x'|} e^{i\vec q.\vec x'} , ~~
F_3(q)=\int d^3x'~\frac{1}{|x'|^3} e^{i\vec q.\vec x'} .
\label{c17}
\end{split}
\end{equation}
Explicit expressions of $F_i(q)$ are calculated in the supplemental material. Final form of  the $k^{th}$ mode of the full wave function (in first order perturbation) is
\begin{equation}
\begin{split}
\psi^{(\vec k)}(&x,t)=\psi^{(k)}_u(x,t)+\psi^{(k)}_s(x,t)\\
=&Ae^{-iE_kt}e^{i( \vec k.\vec x)} + \psi^{(\vec k)}_s(x,t).
\end{split}
\label{psik}
\end{equation}
This is one of our important results. We will use it to  construct the superposition of two waves (with momenta in the same direction) to study QBF.  We consider the wave  just inside horizon, $|x|=\lambda -h(x)$ comprising of  the ingoing modes, as is natural for a BH,  and try to ascertain  QBF outside  horizon, $|x|> \lambda $, as depicted in Fig. \ref{fig:diagram}. The system is reduced to an effectively one (space) dimensional problem, with  QBF observed at  $P$ on the $X$-axis outside the horizon, $x=\lambda $.  

\section {Quantum Back Flow observables} QBF was observed    
\cite{bm} in a superposition of  two plane waves with appropriate mixing coefficients  and  positive momentum to study the probability current $J(x,t)$.  The sign of $J(x,t)$ indicates presence of QBF; for a superposition of negative momenta, the current  will also  be negative, at least  classically. A positive current will indicate the presence of QBF. In our case, it is more convenient to calculate the current directly, following \cite{bm},
\begin{equation}\label{jzt}
J(x,t)=-i\frac{\hbar}{2m}\left(\psi^*(x,t)\frac{\partial \psi(x,t)}{\partial x}-\frac{\partial \psi^*(x,t) }{\partial x}\psi(x,t)\right) 
\end{equation}
using the form of $\psi(x,t) \equiv \psi^{(\vec k)}(x,t)$ we have obtained.

Toy models are a convenient tool in different branches of physics, in particular when a hitherto unexplored effect is being investigated, since for the time being, showing existence of that effect is primary and quantitative estimates for possible vindication can come later. (See for example the work of Berry \cite{ber} where purely plane waves were considered in the study of QBF.) Keeping this in mind, (and following earlier works \cite{bm,years,sol,intro}), let us exploit the toy model of two-mode superposition of plane waves for QBF.  The plane waves are normalized formally as Dirac delta-function normalization or within a finite volume or with the imposition of periodic boundary condition. In existing literature in QBF simple quantum mechanical systems  are considered with explicit wave packets that are localized in a finite volume.   However construction of such a wave packet with finite support in our model is more difficult  (which we plan to pursue in another publication). So in the present work we clearly demonstrate that it is indeed possible to have QBF from a black hole horizon but we refrain from making quantitative predictions. Only difference between our results and  the QBF observables commonly used is that, (since the plane waves do not diminish in magnitude), we will provide the QBF observables  in local form whereas in the latter, spatially integrated expressions are computed. Our framework, in spirit, is similar to the scattering cross-section given per unit area per unit time.

{\bf QBF  from   a  Black Hole event horizon:}
Let us explicitly consider 
\begin{figure}
    \centering
    \includegraphics[width=\columnwidth]{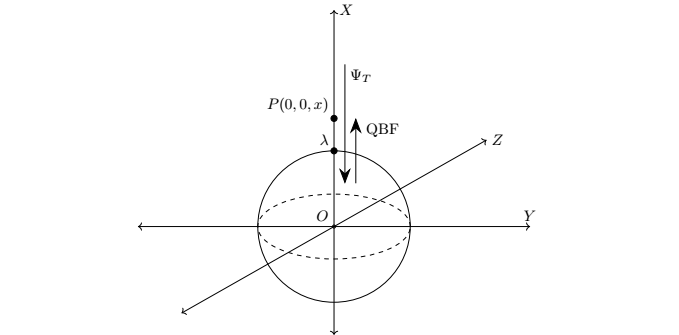}
    \caption{ $P$ denotes the observer and $\lambda$  radius of the event horizon. $\Psi_T(\vec x,t)$ is the superposed wavefunction going into black hole.}
    \label{fig:diagram}
\end{figure}
 $    \psi^{(\vec k)}(\vec x,t) = \psi^{(\vec k)}_u(\vec x,t) + \psi_s^{(\vec k)}(\vec x,t)$
with $\psi_u^{(k)}(x,t) = A \exp\left(i\vec k \cdot \vec x -i E_k t\right)$, with  $\vec{k} = -k \hat{x}$ and $\vec{x} = x\hat{x}$ (position vector of $P$ in Fig. \ref{fig:diagram}), leading to  $\vec{q} = \vec{k} - |\vec{k}|\hat{x} = - 2 k \hat{x}$.

The scattering part is given as,
\begin{equation}\label{psiS}
\begin{split}
\psi^{(k)}_s &= -A \frac{E_k}{2\pi}e^{-i E_k t} \frac{e ^{ik |\vec{x}|}}{|\vec{x}|} \left[E_k (2\pi)^3 \delta(\vec{q}) - E_k \frac{4\pi\lambda}{q^2} +\right.\\
&\left. \frac{2\pi\lambda \hbar^2 c^2}{E_k} k^2 \left(\lim_{\epsilon\rightarrow 0}\partial^2_{q_1} (-Ci(q\epsilon) + \frac{2\sin q\epsilon}{q\epsilon})\right)\right] \\
= &-A \frac{E_k}{2\pi}e^{-i E_k t} \frac{e ^{ik x}}{x} \left[- E_k \frac{4\pi\lambda}{4k^2} + \frac{2\pi\lambda \hbar^2 c^2}{E_k} k^2 \frac{q_1^2}{q^4}\right]
\end{split}
\end{equation}
with  momentum $\vec{q} = q_1 \hat{x}$, $q^2 = q_1^2 + q_2^2 + q_3^2$ and $\lim_{\epsilon\rightarrow 0}\left[\partial^2_{q_1} \left(-Ci(q\epsilon) + \frac{2\sin q\epsilon}{q\epsilon}\right)\right] = -
\frac{q_2^2 + q_3^2 - q_1^2}{q^4}$ 
 simplifies to 
\begin{equation}\label{psiSfin}
\begin{split}
\psi^{(k)}_s (x,&t) = -\frac{A}{2\pi}e^{-i E_k t} \frac{e^{ik x}}{x} \left[-(\hbar^2 c^2 k^2 + m^2 c^4) \frac{\pi\lambda}{k^2} + \frac{\pi\lambda \hbar^2 c^2}{2}\right] \\
&= \frac{A}{2 \pi x} e^{-i(E_k t - k x)} \left[\frac{2GM \pi}{c^2}\left(\frac{\hbar^2 c^2}{2} + \frac{m^2 c^4}{k^2}\right)\right].
\end{split}
\end{equation}
As we have mentioned in the Introduction, there is a universal (numerical value of the) upper bound on the total amount of QBF, but the amount of  QBF is different for different wave functions and one has to come up with an explicit form of wave function that generates a significant amount of QBF. In a simplified toy model approach, we have managed to find  a choice of 
the all-important  two-wave superposition,  $\vec k = - \hat{x}$ and $\vec k = -4\hat{x}$, with an appreciable QBF,
\begin{equation}\label{psiT}
\begin{split}
    \Psi_T (x, &t) = \psi^{(-\hat{x})}(x,t) - 3\psi^{(-4\hat{x})}(x,t)
   \\ 
   &=\left[e^{-i(x - \sqrt{1+m^2}t)} 
     - 3e^{-4i(x - \sqrt{16+m^2}t)}\right]-\\
     \frac{\pi\lambda}{2\pi x}&\left[e^{-i(x - \sqrt{1+m^2}t)}\left(\frac{1}{2}+m^2\right) \right.\\
     &\left. -3 e^{-4i(x - \sqrt{16+m^2}t)}\left(\frac{1}{2}+\frac{m^2}{16}\right)\right]
\end{split}
\end{equation}
where, $\psi^{(\vec k)}(x,t)$ is defined in (\ref{psik}) and we have used the calculated forms of $\psi^{(\vec k)}_u(x,t)$ and $\psi^{(\vec k)}_s(x,t)$. Indeed this not a unique choice and there can very well be better choices of wave functions that might generate larger QBF (within the upper bound). To show that QBF generating superpositions are quite easily available, we have provided two specific examples of different superposition along with their respective QBF, at the end of this section. However, one needs to search for the particular wave function that induces the maximum amount of QBF.

For a general wave superposition of the form
\begin{equation}
\psi(x)=\rho(x)e^{i\int_0^x dx' k(x')}
\label{su1}
\end{equation}
the amplitude function $\rho(x)$ and local wave number function $k(x)$
\begin{equation}
k(x)=\frac{d}{dx}arg~\psi(x)=Im\frac{(d\psi(x)/dx )}{\psi(x)}
\label{su2}
\end{equation}
are real. Positive and negative $k(x)$ will indicate that the wave is travelling in positive or negative direction at $x$. Applying this definition (see for example  \cite{alc,bm,ber} for more details) to a superposition of plane waves with positive momenta (for example)
\begin{equation}
\psi(x)=\sum_{n=0}^{N} C_n e^{ik_nx};~~k_n\geq 0
\label{su3}
\end{equation}
one can show that $k(x)$ can have negative values. As a specific example \cite{alc,bm,ber} for 
\begin{equation}
\psi(x)=1-ae^{ix}
\label{su4}
\end{equation}
the local wave number $k(x)$, given by
\begin{equation}
k(x)=a\frac{a-cosx}{1+a^2-2acosx},
\label{su5}
\end{equation}
will generate QBF for $a<1$ within $|x|<arccos(a)$ periodically.

{\bf Numerical Analysis:} The analytic expressions of the probability and current densities, $\zeta(x,t)=\Psi_T^*(x,t)\Psi_T (x,t)$ and $J(x,t)$ are respectively given by,
	\begin{equation}
	\begin{split}
	&\zeta(x,t)= \frac{1}{1024 x^2}[5(2048 x^2 - 
	64 (16 + 5 m^2) x \lambda +\\
	(&128 + 80 m^2 + 53 m^4) \lambda^2)-
	48(128 x^2 - 4(16 + 17 m^2) x \lambda\\
	+ (&8 + 17 m^2 + 2 m^4) \lambda^2 t)\cos((\sqrt{1 + m^2} - \sqrt{16 + m^2})t - 3 x)],
	\end{split}
	\end{equation}
		\begin{equation}\label{J}
	\begin{split}
	J(x,t) = \frac{\hbar}{(256~m x^2)} [-9472 & x^2 + 4736 x \lambda + 
	832 m^2 x \lambda \\
	- 592 \lambda^2 - 208 m^2  \lambda^2 - 
	73 & m^4 \lambda^2 + 
	30 (128 x^2 - \\
	4 (16 + 17 m^2) x \lambda + (8 + 17 & m^2 + 
	2 m^4) \lambda^2) \cos[(\sqrt{1 + m^2} \\
	- \sqrt{16 + m^2})~t - 
	3 x] + 
	360 & m^2 \lambda \sin[
	\sqrt{1 + m^2}~t \\
	-\sqrt{16 + m^2}~t - 3& x]].
	\end{split}
	\end{equation}

	\begin{figure}
	\centering
	\includegraphics[width=0.75\columnwidth]{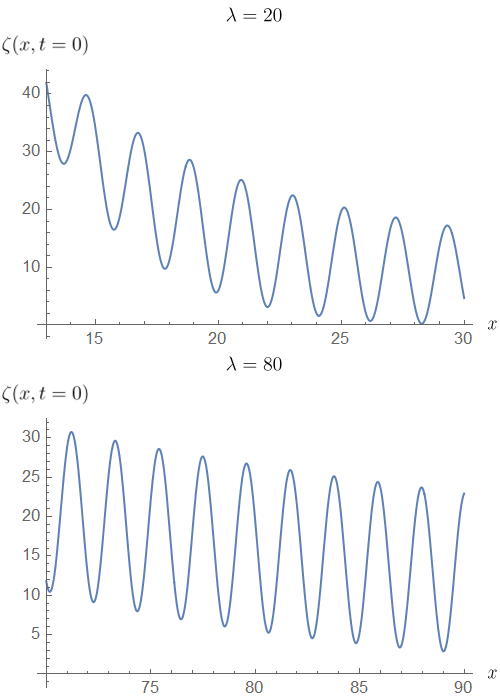} 
	\caption{Plots of the density function $\zeta(x,t=0)$ for two values of $\lambda$.}
	\label{fig:zetaXt0}
\end{figure}
\begin{figure}
    \centering
    \includegraphics[scale=0.7]{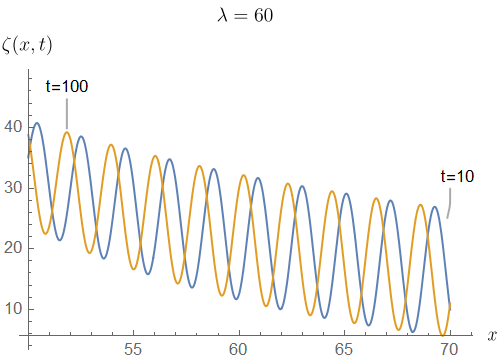}
    \caption{Plot of $\zeta(x,t)$ at two different times, $t=10$ and $t=100$, for $\lambda=60$.}
    \label{fig:zetaXt10t100}
\end{figure}

   The functions are  analytic, $\forall x \neq 0, \forall t \ge 0$, i.e. it is non-analytic only at the curvature singularity at $x=0$ and not at the coordinate singularity at $x=\lambda$. In Fig. \ref {fig:zetaXt0}  we  show the variation in probability density for two values of $\lambda $. There appears to be a qualitative change in the behaviour after the horizon, that is more pronounced for smaller $\lambda $. In Fig. \ref{fig:zetaXt10t100}, snapshots of the density profile are shown for two different times. Notice that the density is greater inside the horizon and as expected, falls off away from the horizon, on the outside. The time-stationary behaviour is readily visible.
\begin{figure}
		\centering
		\includegraphics[width=0.75\columnwidth]{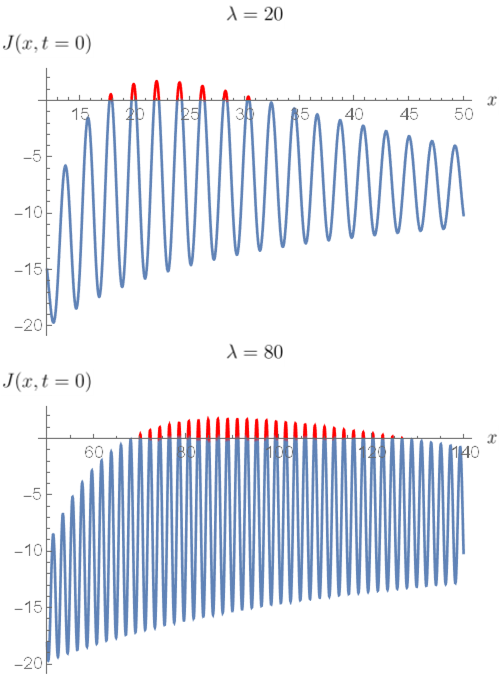}
		\caption{QBF (red region) observed from the plots of $J(x,t=0)$ for two values of $\lambda$, on superposing $\vec k=-\hat{x}$ and $\vec k=-4\hat{x}$ solutions. The size of the QBF region increases with $\lambda$. }
		\label{fig:JXt}
	\end{figure}

\begin{figure}
    \centering
    \includegraphics[scale=0.6]{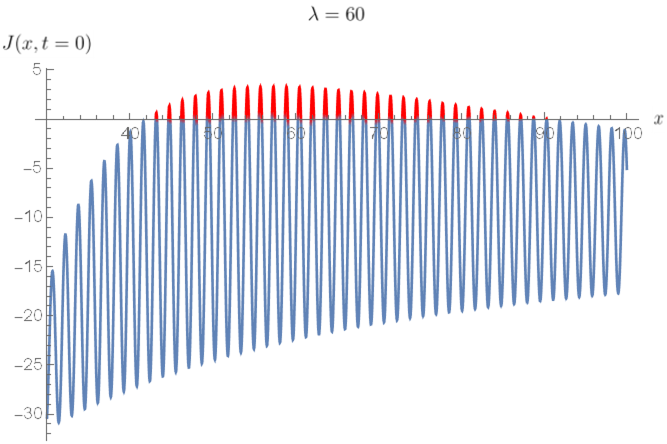}
    \caption{ $J(x,t=0)$ plot showing sizeable QBF on superposing solutions with, $\vec k = -\hat{x}$ and $\vec k = -5\hat{x}$.}
    \label{fig:JXtk15}
\end{figure}
\begin{figure}
    \centering
    \includegraphics[scale=0.6]{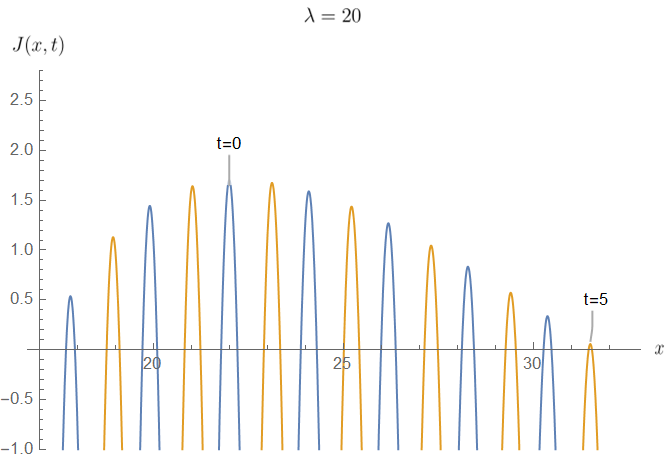}
    \caption{A magnified plot of the QBF region for $t=0$ (blue curve) and $t=5$ (yellow curve), showing evolution of the QBF region with time.}
    \label{fig:peakshift}
\end{figure}
Our most decisive results appear in the profile in Fig. \ref{fig:JXt}, where we plot the current $J(x,t)$ for $\hbar=1, m=3$, for two values of $\lambda$. In Fig. \ref{fig:JXt}, we see that there is a small  QBF region  for $t=0$ (actually for all values of $t$ as well) near the horizon $x=\lambda$, that is shaded red in the figure. It can be shown numerically that $J(x,t)$ is strictly negative (as expected naively) for $x << \lambda$ as well as for $x >> \lambda$. This can also be seen from the time slices in Fig. \ref{fig:JXt}. The envelope of the oscillatory plots in Fig. \ref{fig:JXt} is only positive for a finite region surrounding $x=\lambda$ as shown. This suggests that QBF is {\it originated} from only a finite region around the Schwarzchild BH event horizon. A different superposition and its associated QBF is provided in  Fig. \ref{fig:JXtk15}.

In fact size and shape of the region from where QBF emerges changes (periodically) with time. Such a shift in the QBF peaks is depicted for two different times $t=0$ and $t=5$, in Fig. \ref{fig:peakshift}, an enlarged view of the QBF zone (marked in red) for the $\lambda =20$ case of Fig. \ref{fig:JXt}. Both in Fig. \ref{fig:JXt} and \ref{fig:JXt} it is to be understood that QBF regions occur periodically along the intervals in $x$. In \ref{fig:JXt}  the QBF zone at $t=0$ (marked in blue) are different from the QBF zones at $t=5$ (marked in yellow) indicating how the QBF zones evolve with time. Note that only for plane wave superpositions the corresponding QBF occur periodically.  Similar situations in other QBF models are discussed in detail in \cite{ber}.

\section {\label{sec:rind}Quantum Back Flow for Rindler observers } In this section we will briefly present another example of QBF for a spacetime that has a close connection with the Schwarzschild spacetime but has also essential differences that is reflected in the corresponding QBF; the Rindler spacetime (see for example \cite{rind}). Quite interestingly, even though Rindler metric has a horizon, because of the qualitative difference in the nature of the two horizons the QBF also has a different structure that we will elaborate.

The Rindler horizon appears in the following way. The world line of a point particle moving with constant proper acceleration, (that is the acceleration measured by an observer co-moving with the said point), in flat $(1+1)$-dimensional spacetime, will be a hyperbola. World lines of light rays asymptote to the arms of the hyperbola. Hence the light ray paths constitute horizons bounding the Rindler wedge  in the sense that any signal from outside the wedge can not reach the accelerating particle since to do that it has to overtake the light pulse. On the other hand, from the point of view of the accelerated observer, the light pulse is always at a constant distance ($\sim $ inverse of the proper acceleration) behind. 

Clearly, the Rindler horizon is different from a Schwarzschild black hole horizon: the former will be present so long as the acceleration persists and is not eternal as the latter (in the classical sense, disregarding Hawking radiation). Furthermore the true singular nature of the latter metric, where the curvature is infinite, is absent in the former. Still, the horizon analogy can be pursued further: in the Rindler case, a constant force is required to maintain a  constant proper distance  from the Rindler horizon; in the Schwarzschild case, a  (redshifted) force ($\sim $ surface gravity) on a particle is needed to keep it stationary on the horizon. Again, considering a sufficiently massive black hole, (so that the spacetime curvature is extremely small), similar phenomena (red shift, time dilation and so on)  occur in the vicinity of both the horizons and Unruh radiation in Rindler is an analogue of Hawking radiation in Schwarzschild. With an appropriate rescaling, the Rindler metric structurally reduces to a Schwarzschild form.

A convenient way to study the effects of a constant acceleration is to exploit Rindler coordinates. For acceleration along the direction of motion, it is possible to show that the acceleration is the same for an intertial frame and in the proper frame of the accelerating particle. Hence one can solve the motion of a particle in a Minkowski (laboratory) frame in a straightforward way and transform it to the comoving frame in which the  accelerating  particle is at rest, which in this case is the Rindler frame. Interestingly the metric  in  Rindler frame is that of a curved spacetime, with a horizon, the Rindler horizon. This is agreeable to us since we have already analyzed QBF for a generic curved spacetime and can simply borrow the machinery to compute QBF for Rindler spacetime. However, since the Rindler metric is only defined within the Rindler wedge, it is not feasible to consider QBF across the Rindler horizon. One might heuristically argue that QBF supposedly cannot violate Special Relativity, which would have been the case if QBF across a Rindler horizon was allowed. However, as we demonstrate below, the good news is that QBF is non-vanishing inside the Rindler wedge.

\begin{figure}
	\centering
	\includegraphics[width=0.75\columnwidth]{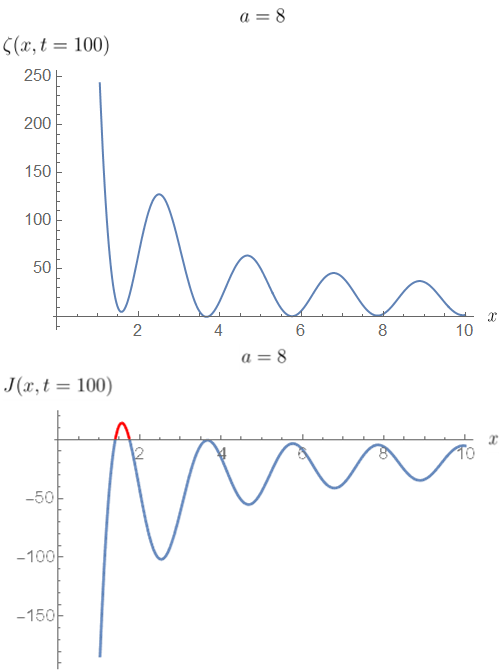} 
	\caption{Plots of the density function $\zeta(x,t=100)$ (top) and the current $J(x,t=100)$ for $|x| \in (\frac{1}{a},10)$.}
	\label{fig:Rin}
\end{figure}

As explained just above, the solution of Newton's dynamical equation, 
\begin{equation}
m a=\frac{dP}{dT}= m\frac{d(\gamma U)}{dT},~~U=dX/dT, ~\gamma =1/({\sqrt{1-U^2/c^2}})
    \label{r1}
\end{equation}
with  $a$,  the proper acceleration of the observer, is given by,
\begin{equation}
U=aT/({\sqrt{1+(aT/c)^2}}).
    \label{r2}
\end{equation}
Defining the proper time $t$ by $dT/dt=1/({\sqrt{1-U^2/c^2}})$ one obtains the transformation laws \cite{rind} 
\begin{equation}\label{XTxt}
\begin{split}
X = \left(x + \frac{1}{a}\right)\sinh[a (t-t_0)] + X_0 - \frac{1}{a},\\
T = \left(x + \frac{1}{a}\right)\cosh[a(t-t_0)] + T_0,
\end{split}
\end{equation}
connecting the flat Minkowski metric
\begin{equation}
ds^2=-dT^2 + dX^2
    \label{r3}
\end{equation}
to the Rindler metric (in terms of proper time and space coordinates $x,t$) given by
\begin{equation}\label{RindMetric}
     ds^2 = -(1+x a)dt^2 + dx^2,
\end{equation}
where we have used the $(-1,+1)$ signature and $c=1$. Clearly, for $x \in \mathbb{R}\setminus \{-1/a\}$, the Minkowski coordinates satisfy, $-T < X < T$ (the Rindler wedge). Following the same procedure outlined in the next two sections, we obtain the scattering part of the wavefunction as, $\psi^{Rind(k)}_s (t,\vec x) = A \frac{E_k}{2 \pi |\vec x|}e^{-i(E_k t - k|\vec x|)}\left[\frac{4 a}{q^2} - 2 \pi \delta(\vec q)\right]$. We construct the superposition of two wavefunctions, with $\vec k = -\hat{x}$ and $\vec k = -4 \hat{x}$ and plot the probability density $\zeta(x,t)$ and current density $J(x,t)$ in Fig. \ref{fig:Rin}.
From the plot of $J(x,t)$ in Fig. \ref{fig:Rin}, we see that QBF indeed occurs (marked in red) away from the Rindler horizon at $|x| = \frac{1}{a}$, i.e. QBF is observed strictly ``inside" the Rindler wedge. Note that there are no divergences in the two profiles of Fig. \ref{fig:Rin} near the horizon; we had to scale it up so that the small amount of QBF shows up clearly, leading to   large but finite  values of the respective  observables near horizon. 
\section {Discussion}
To summarize, we have studied the QBF across the event horizon of a Schwarzschild BH, using  perturbative solutions of the Schrodinger equation for a particle in the BH background, near the horizon. In a superposition of two ingoing modes, QBF is observed across the horizon. Interestingly QBF persists for a finite spatial range. In our  stationary model, overall time dependence and related observations of  charge and current density  are uniform (without any decay). In the present toy model approach,  we have not attempted to provide  quantitative estimates in terms of a conventional observable, the total probability (integrated over position) of QBF. We believe this weakness can be overcome by considering wavepackets, instead of a simple superposition, as done here. However it should be noted that wave propagation and wave packet construction in curved (space and) spacetime is itself an involved problem that has created a lot of interest in recent years as well; some of the novel features related to curved spacetime being decoherence of wave packet due to its interaction with  zero-point fluctuations produced by a gravitational wave, ambiguity in defining a particle excitation, among others (some relevant references are \cite{f1,f2,f3,f4,f5}). Keeping this in mind we have adhered to the toy model presented here which can establish presence of a QBF across BH horizon. We plan to introduce wave packets to study the QBF in this context in near future.

We have also provided a brief description of QBF in Rindler space where it occurs inside the Rindler wedge.

The present work raises several open  questions and possibilities: how do we interpret the QBF across BH horizon? Can the QBF correspond to outgoing modes? Does there exist any form of correlation between the ingoing and outgoing (QBF) part of the wavefunction? Even though QBF is an effect, distinct from  information leakage through Hawking radiation, can it lead to information exchange between inside and outside horizon? Can this type of QBF be realized in analogue gravity models? We conclude by noting several major differences between QBF (in  BH case) and Hawking effect: QBF is a generic quantum  process  present in wave function evolution while, Hawking effect is a semi-classical field theoretic phenomenon applicable near the event horizon of a  BH. QBF is intimately connected to the external matter wave function moving across BH horizon, whereas, no external matter degrees of freedom are involved in  Hawking  effect. Again, there is no  BH mass depletion due to QBF (it can only acquire the mass of the ingoing part of wave function). Also QBF is active for a limited period of time (for wave-packets as ingoing modes), and is restricted in a spatial domain whereas BH mass decreases via Hawking radiation resulting in a BH evaporation. The latter is present throughout the time the BH is alive. 

Since we have mooted the question of a possible connection between QBF and information leakage, it is natural to ask whether there are any comparable time scales involved in the two processes. One such scale for black holes is the Scrambling Time proposed in \cite{sus} that refers to  the process of quick delocalization of quantum information in thermal states. It scales as the logarithm of the number of degrees of freedom of the system. It was also hypothesized that "Black holes are the fastest scramblers in nature" \cite{sus}.  Before proceeding further let us observe that in the Schwarzschild black hole case that we have discussed in detail, we have considered time independent solutions and thus analysis of time scales is beyond the scope of the present work. But, in previous works involving  time dependent QBF \cite{bm,years,intro} it has been shown that generically QBF persists for a finite period of time and it can be non-zero for several time periods. However one should be cautious since the time period can be scaled to different durations with the restriction that the integrated amount of probability transfer by QBF remains the same \cite{bm,years,intro}. Indeed, more study is needed before one can make any conjecture regarding a possible connection between the time scales involved in QBF and information scrambling time (or any other relevant time scale) for black holes.

Open problems: (i) Use  renormalizable ingoing wave packets for quantitative estimates of  QBF.\\
(ii) Investigate QBF in analogue gravity models. As a physically interesting example, in coastal region where a river empties into an ocean, (depending critically upon external parameters), a tidal bore, {\it i.e.} a  strong tidal wave, can occur  that pushes up the river, against the current. In the horizon analogy for a tidal bore, a smooth wavefront is  followed by a series of waves,  travelling upstream forced by high tides \cite{ber2,ber3,berry1}. The bore acts as a horizon, separating an analogue white hole up along the river. There is no wave propagation upstream to the bore and {\it { only rapidly decaying evanescent waves
		can exist there in the analogue white hole region.}} If we push the identification between QBF and tidal bore further then in the white hole situation the leakage of the evanescent waves might be interpreted as the QBF. Recall that QBF also remains active for a finite period of time (provided we use wave-packets for the ingoing modes). Clearly this analogy demands further study.
		
		We end by noting that a deeper analysis of the physical interpretation of the QBF part is needed since apart from an intriguing remark by Berry \cite{rel} regarding possible particle-like nature of the QBF sector, not many observations are present in the literature.\\
	\section{Appendix}
	In the Appendix we have elucidated some parts of our calculations.\\
		\noindent{\bf Equation (11) of main text:} We provide details of the derivation of the solution of Schrodinger equation,  using Green's function  \cite{hertz} 
\begin{equation}
\psi^{(\vec k)}_s(x,t)=\int d^4x'~G_4(t-t'; x-x') V(t',x')\psi^{(\vec k)}_u(t',x')
\label{c10}
\end{equation}
which, can be rewritten as
\begin{equation}
\begin{split}
\psi^{(k)}_s(x,t)&=e^{-iE_kt}\int d^3x'~G_3( x-x') V(x')\psi^{(\vec k)}_u(x') \\
=e^{-iE_kt}&\int d^3x'~\frac{-E_k}{2\pi |x-x'|}e^{ik|x-x'|} \left[-\frac{2 G M}{c^2|x'|}{\cal A}-\right.\\
\hbar^2\nabla^2 \frac{1}{\cal A}&+\frac{2\hbar^2 GM}{|x'|^3}x^{'i}x^{'j}\partial'_i\partial'_j \frac{1}{\cal A} +\left. m^2 c^4\frac{1}{\cal A}\right](Ae^{ik.x'})
\end{split}
\label{c11}
\end{equation}
where, $E_k^2=\hbar^2 c^2 k^2+m^2 c^4$. In the above, $G_4(t-t'; x-x')$ and $G_3( x-x')$ refer to the four and three dimensional Green's functions respectively. We use the notation of \cite{hertz}.  This yields
\begin{equation}
\begin{split}
\psi^{(k)}_s(x,t)&=\frac{-AE_k}{2\pi}e^{-iE_kt}\int d^3x'~\frac{e^{ik|x-x'|}}{|x-x'|}\left[-\frac{\lambda}{|x'|}\sqrt{k^2+m^2} \right.\\
+&\left. k^2
\frac{1}{\sqrt{k^2+m^2}}+m^2\frac{1}{\sqrt {k^2+m^2}}\right.\\
&-\left. \frac{\lambda}{|x'|^3}x^{'i}x^{'j}k_i k _j \frac{1}{\sqrt{k^2+m^2}} \right](e^{ik.x'})
\label{c12}
\end{split}
\end{equation}
After simplification we find,
\begin{equation}
\begin{split}
\psi^{(k)}_s(&x,t)=\frac{-AE_k}{2\pi}e^{-iE_kt}\int d^3x'\\
\frac{e^{ik|x-x'|}}{|x-x'|}&\left[\left(1-\frac{2GM}{c^2|x'|}\right)E_k-\frac{2GM}{E_k}\hbar k_i \hbar k_j \frac{x^{'i}x^{'j}}{|x'|^3} \right](e^{ik.x'}).
\end{split}
\label{c13}
\end{equation}
We concentrate on the region near horizon $|x|=\lambda >>|x'|$ and expand
$$ \rightarrow|x-x'| \approx |x|-\frac{\vec x.\vec x'}{|x|} \equiv |x|\approx \lambda -\hat x .\vec x'$$
so that in  a standard approximation scheme in (\ref{c13}), 
\begin{equation}
\begin{split}
\frac{e^{ik|x-x'|}}{|x-x'|}e^{ik.x'}&\approx \frac{e^{ik(|x|-\hat x.\vec x')}}{|x|-\hat x.\vec x'}e^{ik.x'}\approx \frac{e^{ik|x|}}{|x|}e^{i(\vec k-k\hat x).\vec x'}\\
=&\frac{e^{ik|x|}}{|x|}e^{i\vec q.\vec x'};~\vec q=\vec k-k\hat x .
\end{split}
\label{c14}
\end{equation}	
Substituting these relations in (\ref{c13})  we recover eq.(11) of original text.

\vspace{.5cm}
\noindent{\bf Equation (13) of main text and $F_i(q)$:}
Exploiting spherical symmetry for a generic case yields,
\begin{equation}
F(q)=\frac{4\pi}{q}\int_0^\infty dr~V(r)~ r~ sin(qr).
\label{c18}
\end{equation}
For $V=1/r$, a regularization (in the form of a mass scale $\mu$) is needed;
\begin{equation}
F_2(q;\mu)=\frac{4\pi}{q}\int_0^\infty dr~\frac{e^{-\mu r}}{r}~ r~ sin(qr) =\frac{4\pi}{\mu ^2+q^2}
\label{c19}
\end{equation}
where $\vec q=\vec k-k\hat r;~~q=2k sin(\theta/2)$. Finally taking $\mu \rightarrow 0$ we find
\begin{equation}
F_2(q)=\frac{4\pi}{q^2}.
\label{c20}
\end{equation}
On the other hand, 
for $F_3(q)$ we find, using the same prescription as above, that,
\[F_3(q) = \frac{4\pi}{q} \int_0^\infty \frac{\sin(qr)}{r^2} dr,\]
diverges due to the singularity at $r=0$. Therefore, we perform the integral,
\begin{equation}\label{epsint}
\bar{F}_3(q) = \frac{4\pi}{q} \int_\epsilon^\infty \frac{\sin(qr)}{r^2} dr.
\end{equation}
for some, $\epsilon \in \mathbb{R}^+$.

Then we have,
\begin{equation}\label{ansgen}
\bar{F}_3(q) = 2\pi \left(-\textit{Ci}(q\epsilon)+ 2 \ln(q) - \ln(q^2) + 2 \frac{\sin(q\epsilon)}{q\epsilon}\right),
\end{equation}
where, $\textit{Ci}(z)$ is the Cosine Integral function defined as,
\[\textit{Ci}(z) = - \int_z^\infty \frac{\cos t}{t} dt,\] 
whose series expansion about 0, is given as,
$\textit{Ci}(x) = \gamma + \ln(x) + \sum_{k=1}^\infty \frac{(-x^2)^k}{2k (2k)!}$. Here $\gamma$ is the Euler-Mascheroni constant.

The value of $2 \ln(q)-\ln(q^2)$ is $2 \pi i $ if $q<0$ and $0$ if $q>0$. Thus, more compactly,
\begin{equation}\label{F3ans}
\bar{F}_3(q) = 2\pi\left(-\textit{Ci}(q\epsilon) + 2\pi i \Theta(-q) + 2\frac{\sin(q\epsilon)}{q\epsilon}\right),
\end{equation}
where $\Theta(x)$ denotes the Heaviside-Theta function.
 $F_3(\vec q)$ will be approximated by $\bar{F}_3(\vec q)$ and the limit $\epsilon \rightarrow 0$ will be taken after differentiating $\bar{F}_3(\vec q)$.
 
 -----------------------------------------------------------------
 
1-D result (\ref{psiSfin}) with units: 
\begin{equation}
    \psi_s^{(k)}(x,t) = \frac{A}{2 \pi x} e^{-i(E_k t - k x)} \left[\frac{2GM \pi}{c^2}\left(\frac{1}{2} + \frac{m^2 c^2}{\hbar^2 k^2}\right)\right]
\end{equation}

\section{Acknowledgements} We thank Bibhas Ranjan Majhi for suggestions in the early stage of the work. We are also grateful to Professor Michael Berry and Professor Arseni Goussev for correspondence. We also thank Professor Nava Gaddam and Professor Merab  Gogberashvili for informing us of relevant references. We are thankful to the referees for  constructive comments.


\begin{thebibliography}{99}
\bibitem{berry1} We are thankful to Professor Michael Berry for pointing out the possibility of this analogy.
\end{thebibliography}
\end{document}